\documentstyle[editedvolume,psfig]{crckapb} 

\def\etal{{\it et al. }}
\begin{opening}
\title{GLOBULAR CLUSTERS IN ELLIPTICAL GALAXIES:\protect\\
       CONSTRAINTS ON MERGERS}

\author{DUNCAN A. FORBES}
\institute{School of Physics and Astronomy, \\
	University of Birmingham, Birmingham, B15 2TT, UK}
\end{opening}

\runningtitle{GLOBULAR CLUSTERS}

\begin{document}

% The \begin{document} command comes after the \end{opening}
% command.

\section{Introduction}

There exists a 
relationship between globular cluster mean metallicity and parent galaxy
luminosity (e.g. Brodie \& Huchra 1991; Forbes \etal 1996), which 
appears to be similar to that between 
stellar metallicity and galaxy luminosity. The
globular cluster relation has a similar slope but is 
offset by about 0.5 dex to lower metallicity. The similarity of these
relations suggests that both the globular cluster system and their parent 
galaxy have shared a common chemical enrichment history. If we can
understand the formation and evolution of the globulars, we will also
learn something about galaxy formation. 
With this aim in mind we have created the SAGES (Study of the Astrophysics
of Globular clusters in Extragalactic Systems) project. Project members
include Brodie, Elson, Forbes, Freeman, Grillmair, Huchra, Kissler--Patig
and Schroder. We are using {\it HST} Imaging and Keck spectroscopy to
study extragalactic globular cluster systems. Further details 
can be found at 
http://www.ucolick.org/$^{\sim}$mkissler/Sages/sages.html\\

\section{Results and Discussion}

van den Bergh (1975) has argued that ellipticals have too many 
globular clusters per unit starlight (called specific frequency, S$_N$) 
to be due to the simple merger
of spirals. Spiral galaxies have S$_N$ $\sim$ 0.5. This is increased
to S$_N$ $\sim$ 2 if we take into account the different 
mass-to-light ratios of spirals relative to ellipticals. A typical
elliptical galaxy, with M$_V \sim -21$, has S$_N$ $\sim$ 4. Hence there is
a factor of two difference. This discrepancy gets larger for more luminous
ellipticals. 
However, it has been suggested that new globular clusters can form in the
gas associated with the merger event (Schweizer 1987; Ashman \& Zepf
1992). 
%As indicated above, we require $\sim$ 100\% 
%or more new globulars to
%form compared to the original population of the two spirals. 
Recently proto--globular cluster candidates have been found in a number of
merging systems, largely due to the resolving power of {\it HST}. 
A summary from literature is given in Table 1.

%\begin{quote}
%{\small
%\begin{verbatim}
\begin{table}[htb]
\begin{center}
\caption{Proto--Globular Clusters}
\begin{tabular}{lll}
\hline
Galaxy & Merger Type & $\%$ Increase\\
\hline
NGC 4038/9 & S + S & $\sim$100\\
NGC 7252 & Sc + Sc & $\sim$70\\
NGC 3610 & S + S & $\sim$70\\
NGC 3256 & S + S & $\sim$100\\
NGC 3921 & Sc + S0 & $\sim$40\\
NGC 5128 & E + S & $\sim$20\\
NGC 5018 & E + S & $\sim$10\\ 
NGC 1316 & E + S & 0--30\\
NGC 1275 & E + S & $\le$10\\
\hline
\end{tabular}
\end{center}
\end{table}
%\end{verbatim}}
%\end{quote}

From Table 1 it appears that the number of newly created proto--globulars
varies with the progenitor types (i.e. the gas content) and that
the percentage increase in the total cluster population is 100\% or less. 
This is in contrast to the discussion above which requires 100\% or {\it
more} new globular clusters to counter van den Bergh's objection.  
A further issue to consider, is whether these cluster candidates will
ever resemble globular clusters as we known them -- there is some
circumstantial evidence that they may not (Brodie \etal 1997). 

%Next we examine the globular cluster populations of elliptical galaxies. 
Although the initial evidence was weak, there are 
now a handful of convincing cases 
for bimodal globular cluster color (metallicity)
distributions in ellipticals (see Forbes, Brodie \& Grillmair 1997). 
This indicates that some ellipticals
have more that one globular cluster population. 
%These populations may  
%have formed in distinct star formation episodes from
%different metallicity gas. 
%single star formation episode would give rise to a unimodal distribution. 
The observed metallicity distributions rule out a simple
monolithic collapse and provide a strong constraint for any globular cluster
formation model.  
%In the two cases studied in detail (NGC 4472, Geisler \etal 1996; 
%NGC 1399
%Forbes \etal 1997) it appears that the metal--rich 
%globular clusters have metallicities
%that match 
%the underlying field star metallicity over a wide range in galactocentric
%radius, and they are more centrally concentrated than the metal--poor 
%globular clusters.

Bimodality 
%and radial profiles of the two populations have been taken
has been taken 
as support for the merger picture of Ashman \& Zepf (1992). 
%However when
%the bimodal distributions of several galaxies are examined in detail, it is
%found that they do {\it not} match the expectations of the merger picture
%(see Forbes, Brodie \& Grillmair 1997 for details). In particular, t
Their
model would predict the ratio of metal--rich to metal--poor (N$_R$/N$_P$) 
globulars to be
about 1 for an elliptical with S$_N$ $\sim$ 4 and about 4 for S$_N$ $\sim$
16. Furthermore the metal--poor peak should have a metallicity around
[Fe/H] $\sim$ --1.5, i.e. to match that of a typical spiral. Some examples
of galaxies that do not fit this picture include: NGC 5846, although the
ratio N$_R$/N$_P$ $\sim$ 3, S$_N$ is not high but rather low at 2.8
(Forbes, Brodie \& Huchra 1997). In the case of NGC 4472, the metal--poor
peak contains a total of about 4000 globulars (Geisler \etal 1996). This
would require about 10--20 L$^{\ast}$ spirals. At the other extreme is NGC
3311 (Secker \etal 1995) 
which has virtually no globulars with [Fe/H] = --1.5, i.e. none from
L$^{\ast}$ type spirals. 
 
In Figure 1 
we show S$_N$ against N$_R$/N$_P$ for 12 large ellipticals (4 of which
are cD galaxies). This figure includes two galaxies from the poster of
Geisler \& Lee (this conference). 
If the overabundant globular cluster systems (i.e. those
with high S$_N$ values) are due to the creation of more (metal--rich)
globular clusters in a merger event, then we would expect a trend of
increasing S$_N$ with increasing N$_R$/N$_P$ ratio. Figure 1 shows 
the opposite trend, so that high S$_N$ galaxies actually have a larger
fraction of metal--poor globulars. 

%\begin{quote}
%{\small
%\begin{verbatim}
\begin{figure}
\psfig{figure=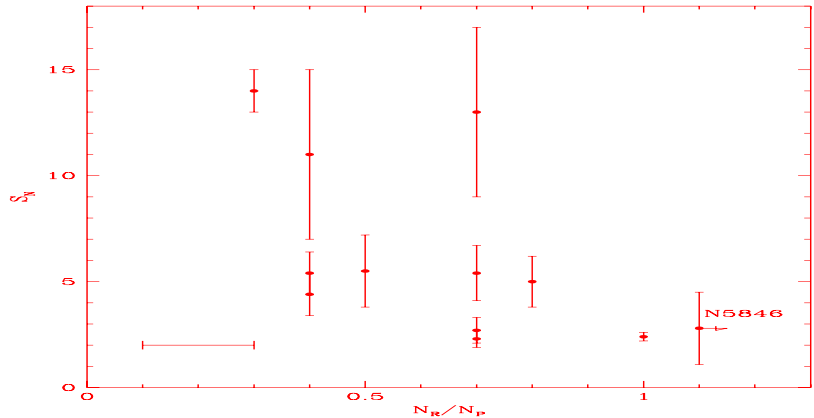,height=6cm,width=4.5in}
%\vspace{5cm}  % amount of vertical space needed
\caption{
The specific frequency (S$_N$) versus the ratio of metal--rich to
metal--poor globular clusters. The merger picture predicts a trend of
increasing S$_N$ with increasing N$_R$/N$_P$ ratio, i.e. 
opposite to that seen.}
\end{figure}
%\end{verbatim}}
%\end{quote}

So although large ellipticals have bimodal metallicity
distributions, when examined in detail they do not fit the expectations of
the merger picture. What is the situation for 
low luminosity ellipticals ? These are much harder to study given that such
galaxies have fewer globular clusters and so any bimodality will be more
difficult to detect. Kissler--Patig, Forbes \& Minitti (1997) have combined
HST and ground--based imaging of NGC 1427 to obtain what is probably the
best photometrically--studied globular cluster system in a 
low luminosity elliptical. 
%For this galaxy we have data on
%$\sim$ 150 globular clusters. This is more than the Milky Way's system and
%represents over 50\% of the galaxy's globular clusters. 
Its color distribution is shown in Figure 2, indicating that it is 
unimodal. If NGC 1427 formed
by the gaseous merger of two disk galaxies we would expect a bimodal
distribution and yet it is clearly unimodal. Unless there is an 
age--metallicity conspiracy that hides the two globular cluster populations
we are forced to conclude that NGC 1427 has only one population and does
not fit the merger picture. 

%\begin{quote}
%{\small
%\begin{verbatim}
\begin{figure}
\psfig{figure=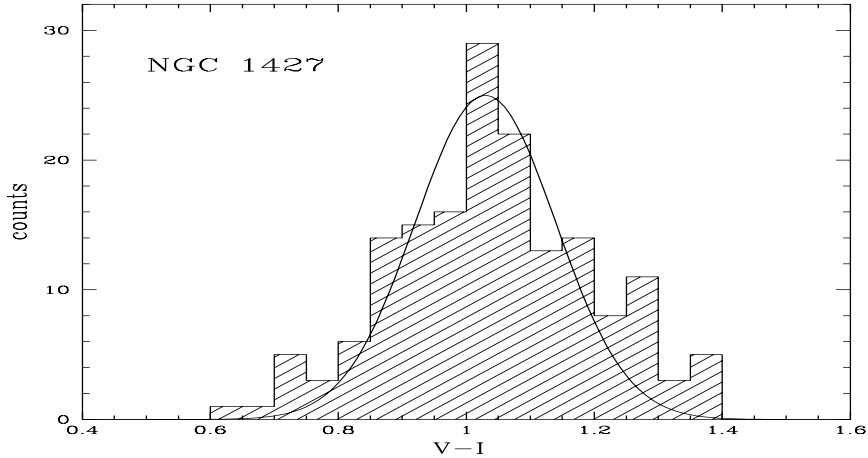,height=6cm,width=4.5in}
%\vspace{5cm}  % amount of vertical space needed
\caption{Globular cluster color distribution 
for NGC 1427 (M$_V$ = --20.5,
S$_N$ = 5). The Gaussian shows the broadening due to photometric
errors. The color distribution is consistent with a single, unimodal 
metallicity. }
\end{figure}
%\end{verbatim}}
%\end{quote}

\section{Concluding Remarks}

It is clear that spiral + spiral mergers are occurring today at redshift z =
0, and that such gaseous mergers may be forming proto--globular clusters
however they may not be forming in sufficient quantities to resemble a
`normal' large elliptical. Gaseous mergers at z $\ge$ 1 predict bimodal
metallicity distributions but these are not seen in small
ellipticals, and the details do not match the predictions for the large
ellipticals. The presence of bimodal distributions rules out a monolithic
collapse, so perhaps we should re--consider a two phase collapse (see
Forbes, Brodie \& Grillmair 1997). 

\section{References}

%\begin{quote}
%{\small
%\begin{verbatim}
%\begin{thebibliography}{}  % Note the empty braces!
%\bibitem{}
Ashman K.M., \& Zepf S.E. 1992, ApJ 384, 50\\
%\bibitem{}
%\bibitem{}
Brodie, J. P., \& Huchra, J. 1991, ApJ, 379, 157\\
%\bibitem{}
Brodie, J. P., \etal 1997, AJ, submitted\\
%\bibitem{}
Forbes, D. A., \etal 1996,
ApJ, 467, 126\\
%\bibitem{}
Forbes D.A., Brodie J.P., \& Grillmair, C.J. 1997, AJ 113, 1652\\
%\bibitem{}
Geisler, D., Lee, M. G., \& Kim, E. 1996, AJ, 111, 1529 \\
%\bibitem{}
Kissler--Patig, M., Forbes, D. A., \& Minitti, D. 1997, MNRAS, submitted\\
%\bibitem{}
Schweizer, F. 1987, Nearly Normal Galaxies, ed. S. Faber (New York:
Springer--Verlag), p 18\\
%\bibitem{}
Secker, J., \etal 1995, AJ, 109, 1019 \\
%\bibitem{}
van den Bergh, S. 1975, ARAA, 13, 217
%\end{verbatim}}
%\end{quote}
%\end{thebibliography}

\end{document}